\documentclass[twocolumn,notitlepage,aps,prb,10pt,superscriptaddress,floatfix,letterpaper,reprint
]{revtex4-2}	

\usepackage{hyperref}
\usepackage[usenames,dvipsnames]{color}
\usepackage{amsmath}
\usepackage[english]{babel}
\usepackage{amssymb}
\usepackage{graphicx}
\usepackage{epstopdf}
\usepackage[normalem]{ulem}
\usepackage{subfigure}
\usepackage{braket}
\usepackage{bbold}
\usepackage{placeins}
\usepackage[para]{threeparttable}
\usepackage{lipsum}
\usepackage{multirow}
\usepackage{bm}
\usepackage{siunitx}
\usepackage{algorithm}
\usepackage{algorithmic}

\usepackage{todonotes}
\usepackage{xcolor}

\usepackage{booktabs}
\usepackage{adjustbox}

\definecolor{linkcolor}{HTML}{399B03}
\definecolor{urlcolor}{HTML}{399B03}
\hypersetup{pdfstartview=FitH, linkcolor=linkcolor,urlcolor=urlcolor,colorlinks=true}
\hypersetup{
    unicode=false,          % non-Latin characters in Acrobat’s  bookmarks
    pdftoolbar=true,        % show Acrobat’s toolbar? 
    pdfmenubar=true,        % show Acrobat’s menu?
    pdffitwindow=false,     % window fit to page when opened
    pdfstartview={FitH},    % fits the width of the page to the window
    pdfauthor={},         % author
    colorlinks=true,        % false: boxed links; true: colored links
    linkcolor=blue,         % color of internal links
    citecolor=red,          % color of links to bibliography
}

\begin{document}

\title{Dynamical Self-energy Mapping (DSEM) for quantum computing}

\author{Diksha Dhawan}%
\affiliation{%
 Department of Chemistry, University of Michigan, Ann Arbor, Michigan 48109, USA
}%
\author{Mekena Metcalf}%
\affiliation{%
Lawrence Berkeley National Laboratory, 1 Cyclotron Rd, Berkeley, CA 94720
}%
\author{Dominika Zgid}%
\affiliation{%
 Department of Chemistry, University of Michigan, Ann Arbor, Michigan 48109, USA
}%
\affiliation{%
 Department of Physics, University of Michigan, Ann Arbor, Michigan 48109, USA
}%

\date{\today}

\begin{abstract}
For noisy intermediate-scale quantum (NISQ) devices only a moderate number of qubits with a limited coherence is available thus enabling only shallow circuits and a few time evolution steps in the currently performed quantum computations. Here, we present how to bypass this challenge in practical molecular chemistry simulations on NISQ devices by employing a classical--quantum hybrid algorithm allowing us to produce a sparse Hamiltonian which contains only $\mathcal{O}(n^2)$ terms in a Gaussian orbital basis when compared to the $\mathcal{O}(n^4)$ terms of a standard Hamiltonian, where $n$ is the number of orbitals in the system. Classical part of this hybrid entails parameterization of the sparse, fictitious Hamiltonian in such a way that it recovers the self-energy of the original molecular system. Quantum machine then uses this fictitious Hamiltonian to calculate the self-energy of the system. We show that the developed hybrid algorithm yields very good total energies for small molecular test cases while reducing the depth of the quantum circuit by at least an order of magnitude when compared with simulations involving a full Hamiltonian.
\end{abstract}

\maketitle
\section{Introduction}
Creating model Hamiltonians is especially important in condensed matter physics when only a few ``important degrees'' of freedom need to be modeled at a more accurate level while the reminder of the physical system of interest can be treated approximately. Such effective Hamiltonians can make intractable physical problems accessible to regular classical computations as well as provide a conceptual understanding of the physical processes present in the system. Effective Hamiltonians are commonly used in embedding methods capable of treating strongly correlated problems~\cite{Georges96,held2007electronic,Kotliar06} where many of such Hamiltonians are recovered as a result of a downfolding procedure. In such a procedure a 1-body Hamiltonian is obtained from a density functional theory (DFT) calculation which is then followed by a projection onto Wannier orbitals~\cite{PhysRev.52.191,RevModPhys.84.1419} and estimation of the lattice model with all the necessary hopping parameters. Subsequently, the 2-body interactions are produced from constrained DFT or random phase approximation (RPA). Effective Hamiltonians can be also obtained as a result of canonical transformation procedure~\cite{yanai2006canonical,yanai2007canonical,chan2007canonical}, L\"owdin orthogonalization method~\cite{sakuma2009effective}, or density matrix downfolding~\cite{zheng2018real,changlani2015density}.
In all these methods, a full, computationally demanding solution of a problem is replaced by a computationally less demanding two step procedure involving the construction of a model Hamiltonian and accurate computation with the resulting Hamiltonian. 

It is straightforward to envision that a variant of such a procedure could be used in quantum chemistry computations involving quantum computers and here in particular calculations on noisy intermediate-scale quantum (NISQ) devices~\cite{preskill2018quantum}. 
Since in quantum computing applications the number of measurements scales with the number of nonzero terms in the Hamiltonian, handling the full Hamiltonian containing $n^4$ 2-body terms, where $n$ is the number of orbitals present in a molecular problem is very challenging particularly for NISQ devices,
where the number of accessible qubits, gates, and circuits depths are very limited.  Consequently, classical--quantum hybrid algorithms resulting in sparse Hamiltonians are naturally best suited computing solutions~\cite{berry2014exponential}.

In this paper, we discuss an idealized classical--quantum hybrid algorithm that we call the dynamical self-energy mapping (DSEM). We believe that this procedure is particularly suitable for handling molecules on NISQ devices.
The classical part of this algorithm is using a polynomially scaling algorithm with respect to the number of orbitals present in the problem in order to produce a total approximate Green's function and self-energy of the molecular problem.  The resulting self-energy is then used to parameterize an effective Hamiltonian. This effective Hamiltonian containing only a small subset of 2-body integrals describes a fictitious system that has the same dynamical part of the self-energy as the parent molecular problem containing all 2-body integrals. Subsequently, this fictitious system described by the sparse Hamiltonian is passed to and solved by the quantum machine that produces its dynamic self-energy. 

Since in this algorithm the quantum machine is dealing only with a very sparse Hamiltonian containing at most $n^2$ 2-body integrals, the number of gates and the circuit depth are severely reduced as compared to a case when all $n^4$ 2-body interactions are present in the molecular Hamiltonian.
Consequently, we expect that the DSEM procedure will be especially relevant for NISQ devices.
Note, that the DSEM algorithm results in a sparse Hamiltonian parameterization for the fictitious system and it gives access to the evaluation of all relevant observables and the electronic energy since it has the same dynamic part of self-energy as the parent system. The ability to rightfully reproduce the self-energy is not always present in  the traditional model Hamiltonians that may be designed to reproduce only specific properties. 

This paper proceeds as follows. In Sec.~\ref{sec:method}, we explain the basics of the DSEM procedure and discuss how the self-energy necessary for the model Hamiltonian evaluation can be approximated. In Sec.~\ref{sec_results}, we demonstrate the accuracy of our procedure. We conclude in Sec.~\ref{sec:conclusions}.

\section{Method}\label{sec:method}

We define a general Hamiltonian for a chemical system of interest as 
\begin{equation}\label{full_ham}
    \hat{H}_{\text full} = \sum_{ij}^{n} t_{ij}a_{i}^{\dagger}a_{j} + \frac{1}{2}\sum_{ijkl}^{n} v_{ijkl}a_{i}^{\dagger}a_{k}^{\dagger}a_{l}a_{j},
\end{equation}
where $v_{ijkl}$, denoted in short as $\langle ij|kl\rangle$, are 2-body Coulomb interactions defined as
\begin{equation}\label{2-body}
    v_{ijkl}=\int\int dr_{1}dr_{2}\phi_{i}^{*}(r_{1})\phi_{j}(r_{1})\frac{1}{r_{12}}\phi_{k}^{*}(r_{2})\phi_{l}(r_{2})
\end{equation}
and containing $n^4$ terms, where $n$ is the number of orbitals present in the full molecular problem.
The 1-body operator is defined as 
\begin{equation}\label{1-body}
    t_{ij}= \int dr_{1}\phi_{i}^{*}(r_{1})h(r_{1})\phi_{j}(r_{1}),
\end{equation}
\begin{equation}
    h(r_{1}) = -\frac{1}{2}\nabla_{2}(r_{1})- \sum_{A} \frac{Z_{A}}{|r_{1}-R_{A}|}.
\end{equation}

For a molecular system of interest the exact Green's function is related to its non-interacting Green's function via Dyson equation
\begin{equation}\label{dyson}
    \Sigma_{\infty} + \Sigma(\omega) = [G_{0}(\omega)]^{-1} - [G(\omega)]^{-1},
\end{equation}
where $\Sigma_\infty$ and $\Sigma(\omega)$ are the static and the dynamical, frequency dependent part of the self-energy, respectively.~\cite{szabo2012modern} Both these self-energies arise due to electronic correlations present in the system of interest.
The zeroth order Green's function is defined as 
\begin{equation}
    G_0(\omega)=[(\omega+\mu)S-F]^{-1},
\end{equation}
where $\mu$, $S$, and $F$ are chemical potential, overlap, and Fock matrix, respectively. Note that here the Fock matrix is defined as 
\begin{equation}
    F_{ij}=t_{ij}+\sum_{kl}\gamma_{kl}(v_{ijkl}-0.5v_{ilkj})
\end{equation}
and can be evaluated using a 1-body density matrix $\gamma$ that does not necessarily need to come from Hartree-Fock but may come from a correlated method.

The first assumption of the DSEM procedure is that for a molecular system, we will be able to produce an approximate self-energy at a low polynomial cost. 
This approximation to the true self-energy is produced in the classical part of the classical-quantum hybrid algorithm and it can be evaluated in multiple ways, for details see Sec.~\ref{sec_approx_self_energy}.
In this paper, we approximated the exact dynamical self-energy either as $\Sigma_1$, the first coefficient of the high frequency expansion~\cite{rusakov2014local} or as $\Sigma^{(2)}(\omega)$, the dynamical second-order self-energy from the second order, finite temperature, fully self-consistent  Green's function method (GF2)~\cite{Dahlen05,Phillips14,Rusakov16,Iskakov19}.
In principle, on a classical machine, the approximate self-energy  can be evaluated using any polynomially scaling algorithm capable of treating a large number of orbitals (eq. GW~\cite{Hedin65}, FLEX~\cite{PhysRevB.55.2122,drchal2004dynamical}, M\o ller-Plesset second order (MP2)~\cite{pople1976theoretical}).

The second assumption of our algorithm is that using this approximate self-energy, a Hamiltonian of the fictitious system, $H_{\rm fic}$ can be evaluated in such a way that with only a subset of 2-body integrals (here at most $n^2$) and all 1-body integrals of the original problem it recovers very well the approximate self-energy of the original molecular problem. This sparse $H_{\rm fic}$ is then used to produce a Green's function and subsequently a self-energy on a quantum machine. Finally, we assume that such a sparse, fictitious Hamiltonian $H_{\rm fic}$ will result in a very shallow quantum circuit requiring only a limited number of qubits due to the sparsity of the 2-body integrals.
Note that such a quantum machine evaluated self-energy is exact for a fictitious, auxiliary system defined using a classical machine, however, it is not an exact self-energy of the Hamiltonian containing all interactions. Nevertheless, we expect that such a self-energy of the fictitious system will approximate the exact self-energy of the true molecular system very well as we will show in the subsequent sections.
The quantum computer evaluated dynamical part of the self-energy can then be used to evaluate the total electronic energy and other desired properties. 

\begin{figure}[bth]
\includegraphics[width=\columnwidth]{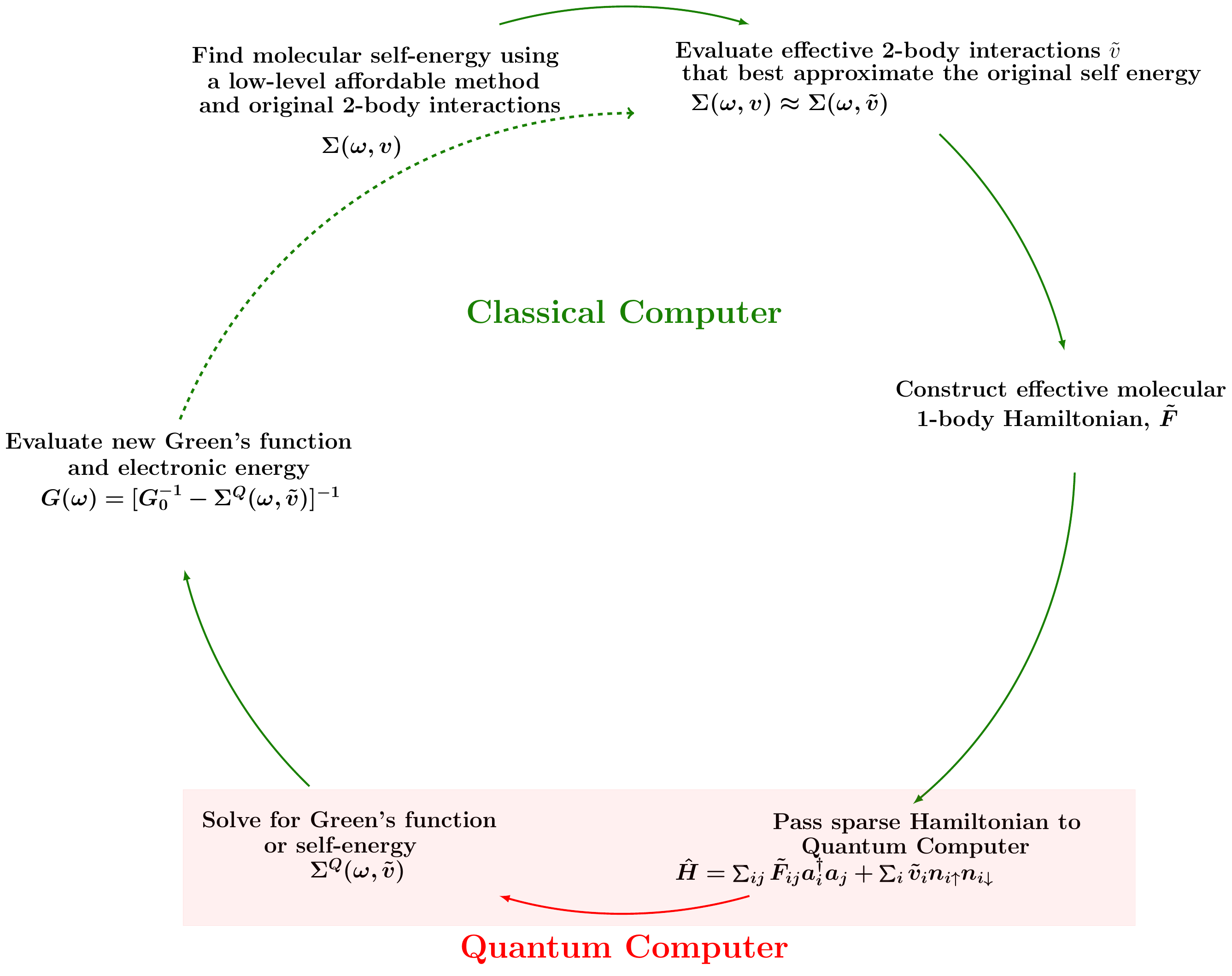}
\caption{A classical--quantum hybrid algorithm using the dynamical self-energy mapping (DSEM) to produce a sparse Hamiltonian for a fictitious system treated by the quantum machine. }
\label{fig:c_q_hybrid}
\end{figure}

Here, we summarize the classical--quantum hybrid algorithm. The {\bf CP} acronym denotes a part of the algorithm executed on a classical machine, the {\bf QP} acronym is used to describe the part executed on a quantum machine. A schematic picture showing the quantum and classical part of the algorithm is presented in Fig.~\ref{fig:c_q_hybrid}.
\begin{description}
\item [CP0] Using the Hamiltonian $\hat{H}_{\text full}$ from Eq.~\ref{full_ham} containing all interactions perform a HF calculation for the system of interest.
\item  [CP1] Employ $\hat{H}_{\text full}$ to evaluate a self-energy defined as $\Sigma_{\infty}(t_{\text full},v_{\text full}) + \Sigma(\omega)(t_{\text full},v_{\text full})$ using a polynomially scaling method (in this paper, we are using GF2). Note, that here we explicitly denote that both parts of the approximate self-energy are evaluated using full 1- and 2-body integrals. 
\item  [CP2] Use the least square fit to find effective integrals $\tilde{v}$ for which $\Sigma(\omega)(\tilde{F},\tilde{v})$ best approximates  $\Sigma(\omega)(t_\text{full},v_\text{full})$ evaluated with full integrals. Note that $\Sigma(\omega)(\tilde{F},\tilde{v})$ is evaluated using fictitious Hamiltonian $H_{\text fic}$ that contains effective 2-body integrals. For details see Sec.~\ref{sec_approx_self_energy} and~\ref{fict_ham}.
\item [QP0] The fictitious Hamiltonian obtained by a classical machine and given by Eq.~\ref{H_fic} 
is passed to the quantum machine.
\item [QP1] Prepare an initial state~\cite{Daniel_Freericks_2020} using several determinants having non-zero overlap with the ground state.
\item [QP2] Evaluate Green's function $G(\omega)(\tilde{F},\tilde{v})$ by using an algorithm described in Refs.~\cite{PhysRevA.64.022319,PhysRevX.6.031045}.
\item [CP3] Evaluate the self-energy  $\Sigma(\omega)(\tilde{F},\tilde{v})$ by using the Dyson equation.
\item [CP4] Employ the quantum machine evaluated self-energy to calculate the new Green's function $G(\omega)$ according to 
\begin{equation}
    G(\omega)=[(\omega+\mu)S-F-\Sigma(\omega)(\tilde{F},\tilde{v})]^{-1}
\end{equation}
Note that by writing $\Sigma(\omega)(\tilde{F},\tilde{v})$, we explicitly denote that this self-energy came from the solution of the fictitious problem on a quantum machine.
\item [CP5] Find chemical potential $\mu$ yielding a proper number of electrons.
\item [CP6] Evaluate a new density matrix $\gamma$ from the Green's function obtained in {\bf CP4} and a new Fock matrix.
\item [CP7] Evaluate 1-body electronic energy as 
\begin{equation}
    E_{\text{1b}}= \frac{1}{2} \sum_{i,j} \gamma_{ij} (t_{ij} + F_{ij}).\\ \label{eq_1b_energy}
\end{equation}
\item [CP8] Using the new Green's function and self-energy evaluate 2-body energy according to
\begin{equation}
E_{\text{2b}}=   \frac{2}{\beta} \sum_{i,j} {\rm Re}[\sum_{\omega}G_{ij}(\omega) \Sigma_{ij}(\omega)(\tilde{F},\tilde{v})].\label{eq_GM}
\end{equation}
\item [CP9] Using the Green's function defined in {\bf CP4}, it is possible to re-evaluate the self-energy on a classical machine and find a new set of effective integrals and continue iterating until electronic energies stop to change
\end{description}

Note that iterations described in {\bf CP9} are optional, in Sec~\ref{sec_results}, we report results without them. Our observations indicated that the difference between performing only the first iteration and all iterations is small (usually below 1 mE$_h$).

\subsection{Approximating self-energy}\label{sec_approx_self_energy}

Here, we focus on possible approximations to the exact self-energy evaluated on a classical machine that can be computed with a polynomial cost. This approximate self-energy is later used to find best effective 2-body integrals for the fictitious Hamiltonian that is used for evaluation of the self-energy on a quantum machine.

\subsubsection{High frequency expansion of the self-energy}\label{sec_hfreq_exp}

In our previous work~\cite{rusakov2014local}, we showed that in certain molecular cases, a good approximation to the exact self-energy is obtained by using a high frequency expansion of the self-energy
\begin{equation}\label{se_expansion}
    \Sigma(\omega) =  \frac{\Sigma_{1}}{\omega} + \frac{\Sigma_{2}}{\omega^2} + \frac{\Sigma_{3}}{\omega^3} + O\big(\frac{1}{\omega^4}\big)
\end{equation}
that is then truncated only to preserve
\begin{equation}
    \Sigma(\omega) \approx \frac{\Sigma_{1}}{\omega}, 
\end{equation}
where $\Sigma_{1}$ is the first coefficients of the high frequency expansion. This coefficient can be evaluated either in approximate perturbative theories or by employing formulas listed in Ref.~\onlinecite{rusakov2014local} that use both 1- and 2-body density matrices. 
Such a simple approximation for the self-energy can be then employed to evaluate fictitious Hamiltonian containing only on-site 2-body integrals given by the following expression
\begin{equation}\label{Ueff}
        \tilde{v}_{iiii} = \sqrt{\frac{2[\Sigma_{1}]_{ii}}{\gamma_{ii}(1 - \frac{1}{2}\gamma_{ii})}},
\end{equation}
where $\gamma_{ii}$ is the on-site 1-body density matrix.

\subsubsection{Frequency dependent self-energy from the second order finite temperature Green's function perturbation theory (GF2)}\label{sec_gf2_se}

For molecular systems, the self-energy obtained in the GF2 method  
\begin{multline}\label{2nd_order_se}
    \Sigma_{ij}^{(2)}(\tau) = -\sum_{klmnpq}G_{kl}(\tau)G_{mn}(\tau)G_{pq}(-\tau) \\
    \times v_{imqk}(2v_{lpnj}-v_{nplj})
\end{multline}
is a very good approximation to the exact self-energy. Therefore, we evaluate it in a classical part of the algorithm using the full molecular Hamiltonian. This evaluation scales as  $n^5n_\tau$, where $n$ is the number of orbitals in the molecular problem while $n_\tau$ is the size of imaginary time grid. 
Since in this approach, all the elements of the self-energy matrix are produced, we use the least square fitting to find a set of sparse 2-body integrals that yield the best approximation to the second order self-energy, namely 
\begin{equation}
    \Sigma_{ij}^{(2)}(\tau, t_{full}, v_{full}) \approx \Sigma_{ij}^{(2)}(\tau,\tilde{F},\tilde{v}). 
\end{equation}
We require the sparse 2-body integrals $\tilde{v}$ depend at most on two indices thus resulting in only $n^2$ 2-body integrals. To check the accuracy of this approximation we tested and included multiple groups of integrals starting from just the on-site integrals and then gradually increasing the set to $n^2$ integrals containing at most two independent indices. 
%such as {\bf (A)} only the on-site integrals 2-body $\langle ii|ii\rangle$, {\bf (B)} on-site and density-density integrals $\langle ij|ij\rangle$, {\bf (C)} on-site, density-density, and exchange integrals $\langle ij|ij\rangle$, $\langle ij|ji\rangle$, {\bf (D)} as well as all $n^2$ integrals (all the groups mentioned above together with cases such as $\langle ij|jj\rangle$).

\subsection{Fictitious Hamiltonian}\label{fict_ham}

Note that using the effective integrals obtained either in Sec.~\ref{sec_hfreq_exp} or \ref{sec_gf2_se}, the sparse Hamiltonian that is used subsequently by the quantum machine has the following form

\begin{equation}\label{H_fic}
    \hat{H}_{\text fic}=\sum_{ij}\tilde{F}_{ij}a^\dagger_i a_j+\frac{1}{2}\sum_{ijkl}\tilde{v}_{ijkl}a^\dagger_i a^\dagger_j a_k a_l,
\end{equation}
where $\tilde{v}_{ijkl}$ are non-zero only for the chosen integrals groups. The modified Fock matrix $\tilde{F}_{ij}$ is given by the following equation
\begin{equation}\label{F_NODC}
    \tilde{F}_{ij} = t_{ij} + \sum_{kl} \gamma_{kl}(v_{ijlk} - \frac{1}{2}v_{iklj}) - \sum_{kl} \gamma_{kl}(\tilde{v}_{ijlk} - \frac{1}{2}\tilde{v}_{iklj}),
\end{equation}
where $F_{ij} = t_{ij} + \sum_{kl} \gamma_{kl}(v_{ijlk} - \frac{1}{2}v_{iklj})$ is the Fock matrix produced in GF2, where $\gamma$ is the 1-body density matrix from GF2 and $v_{ijkl}$ are the full 2-body integrals. The term $\sum_{kl} \gamma_{kl}(\tilde{v}_{ijlk} - \frac{1}{2}\tilde{v}_{iklj})$ corresponds to the double counting correction that should be evaluated with the effective 2-body integrals evaluated as discussed either in Sec~\ref{sec_hfreq_exp} or \ref{sec_gf2_se}.

\section{Results}\label{sec_results}

In this section, we will examine our results from different fictitious Hamiltonian parameterizations, namely
{\bf (p1)} $\langle ii|ii\rangle$ on-site 2-body integrals, {\bf (p2)} both on-site integrals as well as $\langle ii|jj\rangle$, $\langle ij|ij\rangle$  integrals, and {\bf (p3)} on-site integrals and all modified two body integrals with two varying indices out of the total of four  indices, namely $\langle ii|jj\rangle$, $\langle ij|ij\rangle$, and  $\langle ij|jj\rangle$ integrals. Note that while the DSEM scheme is designed to be used as a classical--quantum hybrid algorithm, here, to provide validations and benchmarking of this procedure, we performed it entirely on a regular, classical machine.

\subsection{Hamiltonians with parameterized on-site integrals from high frequency expansion}\label{on-site_ints}
Initially, we parameterized simple molecular Hamiltonians used for the self-energy evaluation to contain  only on-site effective integrals. These integrals can be defined as the 2-body integrals of the form $\tilde{v}_{iiii}=\langle ii|ii\rangle$ where $i$ is the orbital number.  In this paper, to simulate a classical--quantum hybrid computing process, the self-energy was calculated using a polynomially scaling GF2 algorithm.  In this section, we focus on the parameterization of the Hamiltonian using only the on-site 2-body integrals coming from the high frequency expansion of the GF2 self-energy.

In Tab.~\ref{Table_1}, for the H$_{6}$ ring in the STO-6G basis, we list the 2-body integrals obtained from the high frequency expansion of the GF2 self-energy. The 1-body energies and 2-body energies listed were obtained by employing Eq.~\ref{eq_1b_energy} and Eq.~\ref{eq_GM}, respectively. The GF2 and FCI results were evaluated using all $n^4$ integrals, where $n$ is the total number of orbitals in the problem. By FCI($\tilde{v}$), we denote an FCI energy evaluated using effective on-site $\tilde{v}_{iiii}$ 2-body integrals parameterized using the GF2 self-energy.
We note that the total energy that is recovered by FCI($\tilde{v}$) is very close to the true FCI energy and constitutes 102\% of the original correlation energy. The effective on-site integral evaluated in the symmetrized atomic orbital (SAO) basis $\tilde{v}_{iiii}=0.598096$ is smaller than the bare on-site Coulomb integral $v_{iiii}= 0.9060789$ as is expected since it includes the effects of other non-local integrals. 
        \begin{table}
            \centering
            \begin{tabular}{@{}lccc@{}}
                \hline
                 & GF2 & FCI & FCI($\tilde{v}$) \\
                \hline
                2-body integrals & all integrals & all integrals & $\tilde{v}_{iiii}=0.598096$ \\
                
                %\hline
                1-body energy & -9.221596 & -9.162680 & -9.194815\\ % 
                %\hline 
                2-body energy & -0.104371 & -0.185912 & -0.155320\\
                %\hline
                correlation energy & -0.052695 & -0.075320 & -0.076999\\
                 total energy & -9.325966 & -9.348592 & -9.350271  \\
                \hline
            \end{tabular}
            \caption{Energy values obtained using GF2, FCI and parameterized FCI($\tilde{v}$) for H$_{6}$ ring with interatomic distance R=0.95 {\AA} in the STO-6G basis. The second row lists 2-body integrals that were used in the evaluation of self-energies.
            All values of energy are listed in a.u. In case of FCI($\tilde{v}$), $\tilde{v}_{iiii}$ denotes the value of the 2-body on-site integral for $i=1,\dots,6$, all other 2-body integrals are equal to zero.}
            \label{Table_1}
        \end{table}

To assess the effect of the basis set increase, we also performed calculations in the DZ basis. In Tab.~\ref{Table_2}, we list results for H$_6$ ring at R=0.95 {\AA}  in the DZ basis.  Here, the solution of FCI($\tilde{v}$) with parameterized on-site integrals recovers 94\% of correlation energy. Note also the values of the effective integrals $\tilde{v}_{iiii}= 1.001197$  and $\tilde{v}_{jjjj}= 0.424057$ are smaller than the bare on-site Coulumb integrals $v_{iiii}=1.20619$ and $v_{jjjj}= 0.447542$. Such a difference is expected and it is arising due to inclusion of the non-local effects. 
        \begin{table}
        \centering
        \begin{tabular}{@{}lccc@{}}
            \hline
            & GF2 & FCI & FCI($\tilde{v}$) \\
            \hline
            2-body integrals & all integrals & all integrals &  $\tilde{v}_{iiii}$=1.001197\\
            %\hline
            &  &  & $\tilde{v}_{jjjj}$=0.424057\\
            %\hline
            
            1-body energy & -9.257054 & -9.204537 & -9.248869\\ %\hline 
            2-body energy & -0.132880 & -0.206746 & -0.157235\\
            %\hline
            correlation energy & -0.066384  & -0.087733 & -0.082553 \\
            total energy & -9.389935 & -9.411284 & -9.406104 \\
            %\hline
            \hline
        \end{tabular}
        \caption{Energy values obtained using GF2, FCI and parameterized FCI($\tilde{v}$) for H$_{6}$ ring with interatomic distance R=0.95 {\AA} in the DZ basis. The second row lists 2-body integrals that were used in the evaluation of self-energies. 
            All values of energy are listed in a.u. In case of FCI($\tilde{v}$), $\tilde{v}_{iiii}$ denotes the value of the 2-body on-site integral for 1s  $i=1,\dots,6$, $\tilde{v}_{jjjj}$ denotes the value of the 2-body on-site integral for 2s, all other 2-body integrals are equal to zero.}
        \label{Table_2}
    \end{table}
  
\begin{figure*}[bth]
\includegraphics[width=0.47\textwidth]{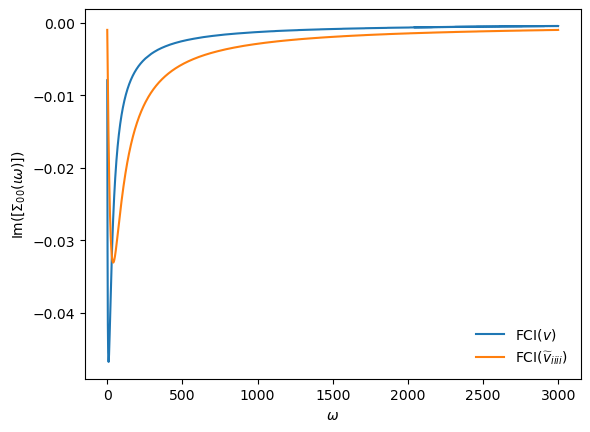}
\includegraphics[width=0.47\textwidth]{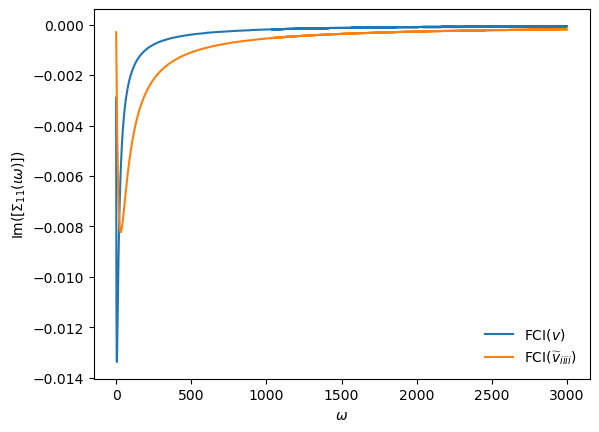}
\includegraphics[width=0.47\textwidth]{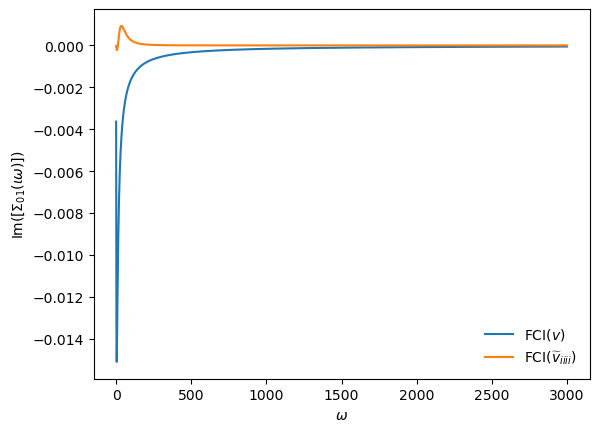}
\includegraphics[width=0.47\textwidth]{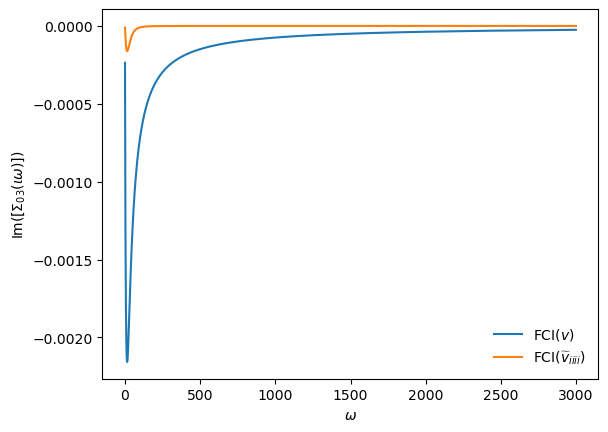}
\caption{The imaginary part of the self-energy for H$_6$ chain in the DZ basis evaluated in FCI with all integrals (denoted here as FCI($v$)) and FCI with effective, on-site 2-body integrals (denoted as FCI($\tilde{v}_{iiii}$)). Top left: Im$[\Sigma(i\omega)]_{00}$ element. Bottom left:  Im$[\Sigma(i\omega)]_{01}$ element. Top right: Im$[\Sigma(i\omega)]_{11}$ element. Bottom right: Im$[\Sigma(i\omega)]_{03}$ element.}
\label{fig_se_fig}
\end{figure*}

While these results are encouraging, in most cases for more complicated molecular examples, employing only on-site 2-body effective integrals cannot lead to the full recovery of the off-diagonal elements of the self-energy.  Since here the off-diagonal elements are only evaluated as a result of  the following multiplication $\Sigma_{ij}^{(2)}(\tau) = -\sum_{ij}[G_{ij}(\tau)]^2 G_{ij}(-\tau)\tilde{v}_{iiii}\tilde{v}_{jjjj}$ not enough freedom may be present to find best on-site integrals $\tilde{v}_{iiii}$ that lead to best approximation $\Sigma_{ij}^{(2)}(\tau,t_{full},v_{full})\approx \Sigma_{ij}^{(2)}(\tau,\tilde{F},\tilde{v}_{iiii})$.
We illustrate this observation in Fig.~\ref{fig_se_fig} by displaying the elements of imaginary part of the self-energy for the H$_6$ chain in the DZ basis.
It is evident that while the diagonal elements of the self-energy are recovered reasonably well, the off-diagonal self-energy elements are not recovered well and are almost equal to zero. Consequently, we conclude that the Hamiltonians with only on-site effective interactions will yield accurate results for systems where the self-energy is majorly diagonal. For other cases a larger number of 2-body integrals is necessary to recover the off-diagonal elements of self-energy.

\subsection{Modified Hamiltonian parameterization using GF2 self-energy}
Here, we focus on parameterization of the fictitious Hamiltonian by finding a small number of effective 2-body integrals that produced a best match between the original GF2 self-energy and the GF2 self-energy evaluated with only the effective $\tilde{v}$ integrals $\Sigma_{ij}^{(2)}(\tau,t_{full},v_{full})\approx \Sigma_{ij}^{(2)}(\tau,\tilde{F},\tilde{v})$ as described in Sec.~\ref{sec_gf2_se}.  In order to perform these fits we used the least-squares optimization subroutine from scipy~\cite{2020SciPy-NMeth}. We investigate parameterizations with both only on-site and more extensive parameterizations with up to $n^2$ integrals.
Here, we perform the DSEM procedure for several small molecular systems such as H$_{6}$ ring, H$_{6}$ chain, H$_{2}$O, and Be dimer. We believe that these systems are good examples of molecular problems that at present can be solved using NISQ devices.

\subsubsection{H$_6$ ring in the STO-6G basis set}

Both the correlation energies and total energies for the H$_6$ ring in the STO-6G basis are presented in Tab.~\ref{tab:H6_ring_p1p2}. We observe that increasing the number of the effective 2-body integrals leads to a significant improvement in the match of the GF2 self-energies $\Sigma_{ij}^{(2)}(\tau,t_{full},v_{full})\approx \Sigma_{ij}^{(2)}(\tau,\tilde{F},\tilde{v})$ and consequently a significant improvement of GF2 total and correlation energies when compared to a GF2 energy evaluated with all the integrals. We analyze three previously mentioned parameterizations p1, p2, and p3. We observed that p2 and p3 parameterizations were completely sufficient to recover the total energy beyond 4th digit after decimal point. 

In Tab.~\ref{tab:H6_ring_p1p2}, we also list the results of FCI (FCI(p1), FCI(p2), and FCI(p2)) performed using the Hamiltonian parameterized at the GF2 level using the three previously discussed parameterizations. We observe that the results from p2 and p3 parameterizations are around 1 mE$_h$ per hydrogen away from the FCI energy evaluated with all the integrals. This loss of accuracy is much smaller than the error currently present on many NISQ devices. Note also that we only applied a relatively naive fitting where we do not use any sophisticated weighing scheme to differently weigh the diagonal and off-diagonal elements of the self-energy or the low- and high frequency behavior.
          \begin{table*}
            \centering
            %\begin{adjustbox}{width=1\textwidth}
            \begin{tabular}{@{}lcccccccc@{}}
                \toprule
                \multicolumn{9}{c}{H$_{6}$ ring; Basis: STO-6G} \\
                \midrule
                 & GF2(v) & GF2(p1) & GF2(p2) & GF2(p3) & FCI(v) &FCI(p1) &FCI(p2) & FCI(p3)\\
                \midrule
%                2-body Integrals & all integrals & param1 & param2 & param1 & param2    \\ 
                Correlation energy & -0.05269 & -0.06202 & -0.05268 & -0.05266 & -0.07532 & -0.06191 & -0.08107 & -0.06694 \\
              \addlinespace[1pt]
                Total energy & -9.32597 &-9.33529 & -9.32595 &  -9.32593 & -9.34859 & -9.33518 & -9.35434 & -9.34021 \\ % 
                \bottomrule
                \end{tabular}
              %  \end{adjustbox}
           \caption{Energy values obtained using GF2, parameterized GF2, FCI, and parameterized FCI. Symbol p1 stands for a parameterization using $\langle ii|ii\rangle$ integrals. Symbol p2 stands for a parameterization using $\langle ii|ii\rangle$, $\langle ij|ij\rangle$, $\langle ij|ji\rangle$ groups of the effective integrals. Symbol p3 stands for a parameterization that uses all the integrals from the p2 group as well as $\langle ij|jj\rangle$   effective integrals. 
            All values of energy are listed in a.u.}
            \label{tab:H6_ring_p1p2}
        \end{table*}                
\begin{table*}[]
\begin{adjustbox}{width=1\textwidth}
\begin{tabular}{lccccccccc}
\toprule
\multicolumn{2}{c}{} & \multicolumn{2}{c}{H$_{6}$ ring; Basis:STO-6G} &\multicolumn{2}{c}{H$_{6}$ chain; Basis: DZ} & \multicolumn{2}{c}{H$_{2}$O; Basis: DZ} & \multicolumn{2}{c}{Be$_{2}$; Basis: 6-31G}  \\
\midrule
& & JW & BK & JW & BK & JW & BK & JW & BK\\
\midrule
\multirow{2}{*}{H(v)} & SQG &  [3.25$\times 10^3$,1.30$\times 10^4$] & [4.23$\times 10^3$,1.72$\times 10^4$] & [1.21$\times 10^5$,2.15$\times 10^5$] & [1.95$\times 10^5$,3.46$\times 10^5$] & [2.34$\times 10^5$,2.79$\times 10^5$] & [3.90$\times 10^5$,4.49$\times 10^5$] & [2.81$\times 10^5$, 2.90$\times 10^5$] & [4.93$\times 10^5$, 10.66$\times 10^5$]\\
                 & CNOT & [5.01$\times 10^3$,1.97$\times 10^4$] & [4.66$\times 10^3$,1.93$\times 10^4$] & [3.12$\times 10^5$,5.54$\times 10^5$] & [2.43$\times 10^5$,5.11$\times 10^5$] & [7.29$\times 10^5$,8.46$\times 10^5$] & [4.98$\times 10^5$,5.91$\times 10^5$] & [10.83$\times 10^5$,10.66$\times 10^5$]  & [6.58$\times 10^5$, 6.72$\times 10^5$]\\
\multirow{2}{*}{H(p1)} & SQG & 3.18$\times 10^2$ & 5.10$\times 10^2$ & 1.35$\times 10^3$ & 2.60$\times 10^3$ & 1.85$\times 10^3$ & 3.65$\times 10^3$ & 1.20$\times 10^3$ & 2.52$\times 10^3$\\
                 & CNOT & 5.72$\times 10^2$ & 5.12$\times 10^2$ & 4.59$\times 10^3$ & 2.96$\times 10^3$ & 7.30$\times 10^3$ & 4.23$\times 10^3$ & 6.39$\times 10^3$ & 2.96$\times 10^3$\\
\multirow{2}{*}{H(p2)} & SQG & 9.18$\times 10^2$ & 8.70$\times 10^2$ & 4.00$\times 10^3$ & 4.19$\times 10^3$ & 5.02$\times 10^3$ & 4.97$\times 10^3$ & 7.24$\times 10^3$ & 6.15$\times 10^3$\\
                 & CNOT & 1.05$\times 10^3$ & 1.14$\times 10^3$ & 6.71$\times 10^3$ & 5.92$\times 10^3$ & 8.56$\times 10^3$ & 7.18$\times 10^3$ & 1.12$\times 10^4$ & 9.91$\times 10^3$\\
\multirow{2}{*}{H(p3)} & SQG & 1.53$\times 10^3$ & 1.85$\times 10^3$ & 6.64$\times 10^3$ & 9.32$\times 10^3$ & 7.70$\times 10^3$ & 1.04$\times 10^4$ & 9.56$\times 10^3$ & 1.11$\times 10^4$\\
                 & CNOT & 2.17$\times 10^3$ & 2.07$\times 10^3$ & 1.59$\times 10^4$ & 1.14$\times 10^4$ & 1.98$\times 10^4$ & 1.31$\times 10^4$ & 2.40$\times 10^4$ & 1.54$\times 10^4$\\
                 \bottomrule
\end{tabular}
\end{adjustbox}
\caption{Number of single qubit (SQG) gates and CNOT gates required to exponentiate the full and fictitious Hamiltonian under various parameterizations for both Jordan-Wigner(JW) and Bravyi-Kitaev(BK) transformations. Symbol p1 stands for a parameterization using $\langle ii|ii\rangle$ integrals. Symbol p2 stands for a parameterization using $\langle ii|ii\rangle$, $\langle ij|ij\rangle$, $\langle ij|ji\rangle$ groups of the effective integrals. Symbol p3 stands for a parameterization that uses all the integrals from the p2 group as well as $\langle ij|jj\rangle$ effective integrals.}
\label{tab:gatecount}
\end{table*}

\subsubsection{H$_6$ chain in the DZ basis set}

We investigated how the accuracy of different parameterizations behaves as the number of orbitals in the basis set is increased. In Tab.~\ref{tab:H6_chain_p1p2}, we list the result of these studies for the H$_6$ chain in the DZ basis. We observe that the GF2 energy evaluated with all the integrals is recovered by parameterization p3 while parameterization p2 is differing only by $\approx$ 3 mE$_h$. When Hamiltonian in parameterization p3 is used in FCI, we observe that the result is approximately 1 mE$_h$ per hydrogen different than the FCI result evaluated with all the integrals. 
          \begin{table*}
            \centering
            \begin{tabular}{@{}lcccccccc@{}}
                \toprule
                \multicolumn{9}{c}{H$_{6}$ chain; Basis: DZ} \\
                \midrule
                 & GF2(v) & GF2(p1) & GF2(p2) & GF2(p3) & FCI(v) &FCI(p1) &FCI(p2)& FCI(p3)\\
                \midrule
%               2-body Integrals & all integrals & $U_{i}$ =  \\ 
                \addlinespace[1pt]
                Correlation energy & -0.06677 & -0.08606 &  -0.06958 & -0.06623 & -0.09584 & -0.05678 & -0.08102 & -0.10139\\
                \addlinespace[1pt]
                Total energy & -8.13663 & -8.15592 & -8.13944 & -8.13609& -8.16570 & -8.12664 & -8.15088 & -8.17125\\ % 
                \bottomrule
        \end{tabular}
           \caption{Energy values obtained using GF2, parameterized GF2, FCI, and parameterized FCI. Symbol p1 stands for a parameterization using $\langle ii|ii\rangle$ integrals. Symbol p2 stands for a parameterization using $\langle ii|ii\rangle$, $\langle ij|ij\rangle$, $\langle ij|ji\rangle$ groups of the effective integrals. Symbol p3 stands for a parameterization that uses all the integrals from the p2 group as well as $\langle ij|jj\rangle$   effective integrals. 
            All values of energy are listed in a.u.}
            \label{tab:H6_chain_p1p2}
        \end{table*}

\subsubsection{H$_2$O in DZ basis set}

Studying both the H$_6$ chain and ring examples in two different basis sets allow us to confirm that when the p3 parameterization in GF2 is very close to the original GF2 energy evaluated with all the integrals then the FCI energies recovered from the p3 parameterization is also very close to the original FCI energy. This observation prompts us to analyze examples where
calculating the Green's function in the FCI procedure will result in a significant computational time and memory use. For these cases, we will only examine the systematic improvement present in the p1, p2, and p3 GF2 parameterizations.

In Tab.~\ref{tab:H2O_p1p2}, we list GF2 energies resulting from the different parameterizations of the Hamiltonian. Note that both the p2 and p3 parameterizations are only 6 and 2 mE$_h$ away from the original GF2 energy, respectively. 
          \begin{table}
            \centering
           % \begin{tabular}{@{}lcccccccc@{}}
                        \begin{tabular}{@{}lcccc@{}}
                \toprule
%                \multicolumn{9}{c}{H$_{2}$O; Basis: DZ} \\
               \multicolumn{5}{c}{H$_{2}$O; Basis: DZ} \\
                \midrule
                 & GF2(v) & GF2(p1) & GF2(p2) & GF2(p3) \\%& FCI(v) &FCI(p1) &FCI(p2) & FCI(p3)\\
                \midrule
%                2-body Integrals & all integrals & $U_{i}$ =  \\ 
                \addlinespace[1pt]
                Correlation energy & -0.13287 & -0.19069 &  -0.13844& -0.13519 \\ %&  & &\\
                                \addlinespace[1pt]
                Total energy(a.u.) & -85.47654 & -85.53436 & -85.48212 & -85.47886 \\ %& & \\ % 
                \bottomrule
           \end{tabular}
            \caption{Energy values obtained using GF2, and parameterized GF2. Symbol p1 stands for a parameterization using $\langle ii|ii\rangle$ integrals. Symbol p2 stands for a parameterization using $\langle ii|ii\rangle$, $\langle ij|ij\rangle$, $\langle ij|ji\rangle$ groups of the effective integrals. Symbol p3 stands for a parameterization that uses all the integrals from the p2 group as well as $\langle ij|jj\rangle$ effective integrals. 
            All values of energy are listed in a.u.}
            \label{tab:H2O_p1p2}
        \end{table}

\subsubsection{Be$_{2}$ in 6-31G basis set}

Finally, in Tab.~\ref{tab:Be_p1p2}, we analyze a small diatomic molecule Be$_2$ that when calculated in the 6-31G basis set is an ideal test case for calculations on the NISQ devices. For this case, similar to the previous cases the GF2 energy coming from the p1 parametrization is not acceptable. However, the p2 parameterization results in energies that are very close to the original GF2 energy. The p3 parameterization yields the energy value with an error as small as 1 mE$_h$.
          \begin{table}
            \centering
            \begin{tabular}{@{}lcccc@{}}
                \toprule
                \multicolumn{4}{c}{Be$_{2}$; Basis: 6-31G} \\
                \midrule
                 & GF2(v) & GF2(p1) & GF2(p2) & GF2(p3) \\
                \midrule
%                2-body Integrals & all integrals & $U_{i}$ =  \\ 
                Correlation energy & -0.04814 & -0.02047 & -0.04463& -0.04651\\
             \addlinespace[1pt]
                Total energy &  -29.18194 & -29.15427 & -29.17843 & -29.18031\\ % 
                \bottomrule
           \end{tabular}
            \caption{Energy values obtained using GF2, and parameterized GF2. Symbol p1 stands for a parameterization using $\langle ii|ii\rangle$ integrals. Symbol p2 stands for a parameterization using $\langle ii|ii\rangle$, $\langle ij|ij\rangle$, $\langle ij|ji\rangle$ groups of the effective integrals. Symbol p3 stands for a parameterization that uses all the integrals from the p2 group as well as $\langle ij|jj\rangle$ effective integrals. 
            All values of energy are listed in a.u.}
            \label{tab:Be_p1p2}
        \end{table}

\subsection{Number of gates for different Hamiltonian parameterizations} 
   To illustrate that such simplified Hamiltonian parameterizations will lead to a low circuit depth, we calculated the number of gates for the fictitious Hamiltonians constructed in the previous section. Number of gates required for exponentiation of full and fictitious Hamiltonians are listed in Tab.~\ref{tab:gatecount}. We used Jordan-Wigner~\cite{jordan1928pauli} and Bravyi-Kitaev~\cite{Bravyi_2002} transformations for expressing the molecular Hamiltonian in terms of Pauli operators. These transformations were obtained using OpenFermion's~\cite{mcclean2017openfermion} practical implementation of these techniques. Number of gates required were calculated using these Pauli representations as described in Ref.~\onlinecite{Seeley2012}. 
   For the parameterized cases (H(p1), H(p2), H(p3)), we have worked in SAO basis, while for the full Hamiltonian, we evaluated the number of gates both in SAO and molecular orbitals (MO) representation listing both cases.
   For the small molecular cases analyzed here, we observe a reduction in the number of gates by about an order of magnitude when using the fictitious Hamiltonian (even within p3 parameterization) in comparison to the full Hamiltonian. Note, however, that as the system size increases, we expect that the difference in the number of gates necessary to perform a single Trotter step will increase for the full Hamiltonian. Consequently, in the limit of a large molecular system, a parameterized Hamiltonian will result in even larger reduction of necessary gates when compared to the full Hamiltonian.

\section{Conclusions}\label{sec:conclusions}
We have presented a DSEM procedure which allows us to find a fictitious, sparse Hamiltonian that recovers the self-energy evaluated with the full, original Hamiltonian containing all 2-body integrals. DSEM procedure is a classical--quantum hybrid algorithm which employs a classical machine to perform polynomially scaling evaluation of the self-energy that is necessary for finding a sparse, fictitious Hamiltonian that is then used to evaluate an exact self-energy using a quantum machine. Since the quantum machine deals only with the sparse Hamiltonian containing at most $n^2$ terms, the resulting circuit is shallow with many fewer gates than for the circuit necessary to represent the original Hamiltonian. 

We have compared the GF2 and FCI energies obtained using the DSEM procedure to the original GF2 and FCI energies. From these comparisons, we have demonstrated that the errors can be controlled and are small in comparison to the errors currently present for the NISQ machines. Moreover, we have shown that the number of necessary gates in order to express the sparse Hamiltonian is at least one order of magnitude smaller, even for small systems, when compared to the gates necessary to express the original Hamiltonian. 

There are various other quantum algorithms that leverage the sparsity of the Hamiltonian for a reduced gate count in a quantum circuit. In Ref.~\onlinecite{babbush2018low}, Babbush et al. reduced the number of Hamiltonian terms to $\mathcal{O}(N^{2})$, where $N$ is the number of basis functions in a plane-wave basis-set . They diagonalize different components of the Hamiltonian operator, namely kinetic operator and potential operator using plane wave and dual plane wave basis respectively. Our method differs from this approach as we perform all our calculations in a Gaussian orbital basis, which requires fewer basis functions to obtain the same level of convergence with respect to the basis set. For regular chemical systems the number of necessary plane waves are thousands times larger than the number of Gaussian orbitals, $N>>n$. However, at the same time, we make approximations to the self-energy in order to reduce the number of terms in the Hamiltonian which when not done carefully can lead to a loss of accuracy. Further work to improve these approximations is in progress in our lab.
  
We believe that our work opens several new venues of molecular quantum computing research that were not explored before. First, we propose that the quantum machine performs an evaluation for the fictitious Hamiltonian that is used to recover the molecular frequency dependent self-energy. From a mathematical view point, this has several interesting implications. For the ``true'', analytical, and exact self-energy there is only one Hamiltonian capable of yielding this self-energy. However, for a  self-energy that is numerical and only agrees to a very good numerical accuracy with the ``true'' exact self-energy, there can be several Hamiltonians reproducing it. These Hamiltonians can be much more suitable for quantum computing than the original Hamiltonian. Additionally, the analytical properties of Green's functions and self-energies such as the high frequency expansion are well known, thus providing an additional tool in the assessment of the errors arising in computation on NISQ devices. Moreover, these errors can be partially corrected since the analytic limits are known when classical--quantum hybrid algorithms such as DSEM are employed. Finally, we would like to mention that DSEM can be naturally extended to work in conjunction with an embedding framework such as dynamical mean field theory (DMFT)~\cite{georges1992hubbard,Georges96,PhysRevX.6.031045,Kotliar06} or self-energy embedding theory (SEET)~\cite{Kananenka15,Zgid17,doi:10.1021/acs.jctc.8b00927,Iskakov20}.
\section{Acknowledgements}
D.Z. and  D.D. acknowledge National Science Foundation grant number CHE1836497. M. M. was supported by the ``Embedding Quantum
Computing into Many-body Frameworks for Strongly Correlated Molecular and Materials Systems" project, which is
funded by the U.S. Department of Energy(DOE), Office of
Science, Office of Basic Energy Sciences, the Division of Chemical Sciences, Geosciences, and Biosciences.
D.Z. thanks James Freericks for helpful discussions.

\end{document}